\documentclass[natbib,final,5p,times,twocolumn,number]{elsarticle}
\usepackage{amssymb}
\usepackage{lipsum}
\usepackage{textcomp}
\usepackage{amsmath}
\usepackage{float}
\usepackage{ulem}
\usepackage{hyperref}
\usepackage{graphicx}

\usepackage{xcolor} 
\newcommand{\highlight}[1]{\textcolor{black}{#1}}

\journal{Physics Letters B}

\begin{document}

\begin{frontmatter}

\title{Precise measurement of the $\gamma$-decay probability of the Hoyle state \\with a new triple coincidence-detection method}

\author[1]{K.~Sakanashi\corref{cor1}}
\ead{sakanashi@ne.phys.sci.osaka-u.ac.jp}
\cortext[cor1]{First author}

\author[1]{T.~Kawabata\corref{cor2}}
\ead{kawabata@phys.sci.osaka-u.ac.jp}
\cortext[cor2]{Corresponding author}

\author[4]{S.~Adachi}
\author[5]{H.~Akimune}
\author[8]{S.~Aogaki}
\author[8]{D.L.~Balabanski}
\author[8]{S.~R.~Ban}
\author[9]{R.~Borcea}
\author[9]{Ș.~Călinescu}
\author[9]{C.~Clisu}
\author[8]{R.~Corbu}
\author[9]{C.~Costache}
\author[8]{A.~Covali}
\author[8]{M.~Cuciuc}
\author[8]{A.~Dhal}
\author[9]{I.~Dinescu}
\author[9]{N.~Florea}
\author[1,13]{T.~Furuno}
\author[9]{I.~Gheorghe}
\author[9]{A.~Ionescu}
\author[4]{M.~Itoh}
\author[6,7]{S.~Kubono}
\author[8,10]{A.~Ku\c{s}o\u{g}lu}
\author[5]{Y.~Matsuda}
\author[9]{C.~Mihai}
\author[9,11]{R.~E.~Mihai}
\author[9]{C.~Neac\c{s} su}
\author[8]{D.~Nichita}
\author[2]{R.~Niina}
\author[3]{S.~Okamoto}
\author[8]{H.~Pai}
\author[8]{T.~Petruse}
\author[12]{M.~Sferrazza}
\author[8]{O.~Sîrbu}
\author[8]{P-A.~S$\rm\ddot{o}$derstr$\rm\ddot{o}$m}
\author[8]{A.~Spătaru}
\author[9]{L.~Stan}
\author[2]{A.~Tamii}
\author[8]{D.A.~Testov}
\author[9]{A.~Turturica}
\author[8]{G.~Turturică}
\author[9]{S.~Ujeniuc}
\author[8]{V.~Vasilca}

\affiliation[1]{
            organization={Department of Physics, Osaka University},
            addressline={Toyonaka, Osaka 560-0043, Japan}
}
\affiliation[4]{
organization={Research Center for Accelerator and Radioisotope Science, Tohoku University},
addressline={Aoba, Sendai, Miyagi 980-8578, Japan}
}
\affiliation[5]{
organization={Department of Physics, Konan University},
addressline={Kobe, Hyogo 658-8501, Japan}
}

\affiliation[8]{
organization={Extreme Light Infrastructure - Nuclear Physics, ``Horia Hulubei''
National Institute for R$\&$D in Physics and Nuclear Engineering},
addressline={\\30 Reactorului Street, 077125 Magurele, Romania}
}

\affiliation[9]{
organization={``Horia Hulubei'' National Institute for R{$\&$}D in Physics and Nuclear Engineering},
addressline={30 Reactorului Street, 077125 Magurele, Romania}
}

\affiliation[13]{
organization={Department of Applied Physics, University of Fukui},
addressline={Fukui, Fukui 910-8507, Japan}
}

\affiliation[6]{
organization={Nishina Center for Accelerator-Based Science, RIKEN},
addressline={Wako, Saitama 351-0198, Japan}
}

\affiliation[7]{
organization={Center for Nuclear Study, University of Tokyo},
addressline={Wako, Saitama 351-0198, Japan}
}

\affiliation[10]{
organization={Department of Physics, Faculty of Science, Istanbul University},
addressline={Vezneciler/Fatih, 34134, Istanbul, Turkey}
}
\affiliation[2]{
organization={Research center for Nuclear Physics, Osaka University},
 addressline={Ibaraki, Osaka 567-0047, Japan}
}
\affiliation[3]{
organization={Department of Physics, Kyoto University},
addressline={Sakyo, Kyoto 606-8502, Japan}
}
\affiliation[12]{
organization={Départment de Physique, Université Libre de Bruxelles},
addressline={Bruxelles 1050, Belgium}
}

\affiliation[11]{
organization={Institute of Experimental and Applied Physics,
Czech Technical University in Prague},
addressline={Husova 240/5 110 00, Czech Republic}
}

\begin{abstract}
We measured the $\gamma$-decay probability of the Hoyle state with a new method of triple coincidence detection of a scattered $\alpha$ particle, a recoil $\rm ^{12}C$ nucleus, and a $\gamma$ ray in inelastic alpha scattering on $\rm ^{12}C$.
This method successfully enabled a low-background measurement and a precise determination of the $\gamma$-decay probability of the Hoyle state as $\Gamma_\mathrm{\gamma}/\Gamma=[4.00 \pm 0.22 \mathrm{(sta.)} \pm 0.18 \mathrm{(sys.)}]\times10^{-4}$, which is consistent with the previous literature value.
Therefore, we concluded that the literature value can be reliably used in the study of nucleosynthesis in the universe.

\end{abstract}

\begin{keyword}

Triple alpha reaction \sep Hoyle state \sep $\gamma$-decay probability \sep Nucleosynthesis

\end{keyword}

\end{frontmatter}

\section{Introduction}
\label{introduction}

The triple alpha (3$\alpha$) reaction is one of the most important reactions in nucleosynthesis, because it is the doorway reaction that bypasses the $A$ = 5 and 8 bottlenecks and leads to the production of heavier nuclei~\cite{Burbidge1957}.
In the 3$\alpha$ reaction, an $\alpha$ particle is captured by a 2$\alpha$ resonant state in $\rm ^{8}Be$, and a 3$\alpha$ resonant state in $\rm ^{12}C$ is formed~\cite{Hoyle:1953}.
Most of the 3$\alpha$ states decay back to 3$\alpha$, however, they undergo radiative decay to the ground state by emitting $\rm \gamma$ rays or an $\rm e^{+}e^{-}$ pair with a small probability.
At normal stellar temperatures around $T \sim 10^{8}$ K, the 3$\alpha$ reaction via the $0^+_2$ state at $E_{x} = 7.65$ MeV in $^{12}$C, called Hoyle state, is dominant.
Therefore, the radiative-decay probability of the Hoyle state is a very important parameter to determine the 3$\alpha$ reaction rate in the nucleosynthesis.

The radiative-decay probability was extensively measured in the 1960s and 1970s with two methods: a $^{12}$C detection method~\cite{Seeger:1963,Hall:1964,Chamberlin:1974,Davids:1975,Mak:1975,Markham:1976} and a $\gamma$ detection method~\cite{Alburger:1961,Obst:1976}.
In the $^{12}$C detection method, the radiative-decay probability was directly measured by detecting the $^{12}$C ground state from the remnant of the radiative de-exitation of the Hoyle state.
On the other hand, in the $\gamma$ detection method, the $\gamma$-decay probability was measured by detecting the $\gamma$ rays emitted from the Hoyle state, and the radiative-decay probability was determined by combining it with the $\rm e^{+}e^{-}$ pair-decay probability measured in other experiments~\cite{Obst:1972,Robertson:1977,Alburger:1977}.
Most of the results except Ref.$\,$\cite{Seeger:1963} are consistent with each other within their uncertainties, and $\Gamma_{\rm rad} / \Gamma$ = 4.16(11) $\times$ $10^{-4}$ is recommended as the averaged value by the experimental compilation~\cite{Kelley:2017}.
Recently, the radiative-decay probability was re-measured by Kib$\rm \acute{e}$di $et$ $al.$ by detecting the two $\gamma$ rays from the cascade decay of the Hoyle state populated by the $\rm ^{12}C(p,p^{\prime})$ reaction~\cite{kibedi2020}.
Kib$\rm \acute{e}$di $et$ $al.$ reported a  new value of $\Gamma_{\rm rad} / \Gamma$ = 6.2(6) $\times$ $10^{-4}$ which is 50$\%$ higher than the recommended value in Ref.~\cite{Kelley:2017}.

Since the report from Kib$\rm \acute{e}$di $et$ $al.$, the measurement of the radiative-decay width of the Hoyle state has been actively conducted. 
The first to report the radiative-decay width of the Hoyle state after Kib$\rm \acute{e}$di $et$ $al.$ was Tsumura $et$ $al.$ who measured the surviving $^{12}$C~\cite{tsumura}, and reported a value closer to the recommended value than that of Kib$\rm \acute{e}$di $et$ $al.$
Subsequently, Luo $et$ $al.$~\cite{newpaper2}, Dell'Aquila $et$ $al.$~\cite{newpaper3}, and Rana $et$ $al.$~\cite{newpaper4} performed measurements of the surviving $^{12}$C, and their results were all consistent with the recommended value.
Rana $et$ $al.$~\cite{newpaper4} and Paulsen $et$ $al.$~\cite{Paulsen:2025} measured the $\gamma$ rays, following the approach of Kib$\rm \acute{e}$di $et$ $al.$, and their results also supported the recommended value.
Moreover, Paulsen $et$ $al.$ reanalyzed the data of Kib$\rm \acute{e}$di $et$ $al.$~\cite{kibedi2020}, and demonstrated that these data could be interpreted as consistent with the recommended values.

Although numerous experiments have been conducted, they only measured either $^{12}$C or $\gamma$ rays.
Simultaneous detection of both $^{12}$C and $\gamma$ rays could reduce background events and lead to a precise measurement. 
However, measuring both $^{12}$C and $\gamma$ rays simultaneously is challenging. 
Since the detection efficiency of $\gamma$ rays is low, a thick target is needed to accumulate sufficient statistics, but a thicker target makes the detection of $^{12}$C difficult.
To date, only one experiment that measured both $^{12}$C and $\gamma$ rays simultaneously has been published, and it reported a radiative-decay probability about 4.5 times larger than the recommended value~\cite{newpaper1}.

In this study, we aim to precisely measure the $\gamma$-decay probability of the Hoyle state by simultaneously detecting $^{12}$C and $\gamma$ rays using a Si strip detector and a large solid-angle LaBr$_3$(Ce) and CeBr$_3$ array.

\section{Experimental procedure}
\begin{figure}[htb]
\centering
\includegraphics[width=8cm]{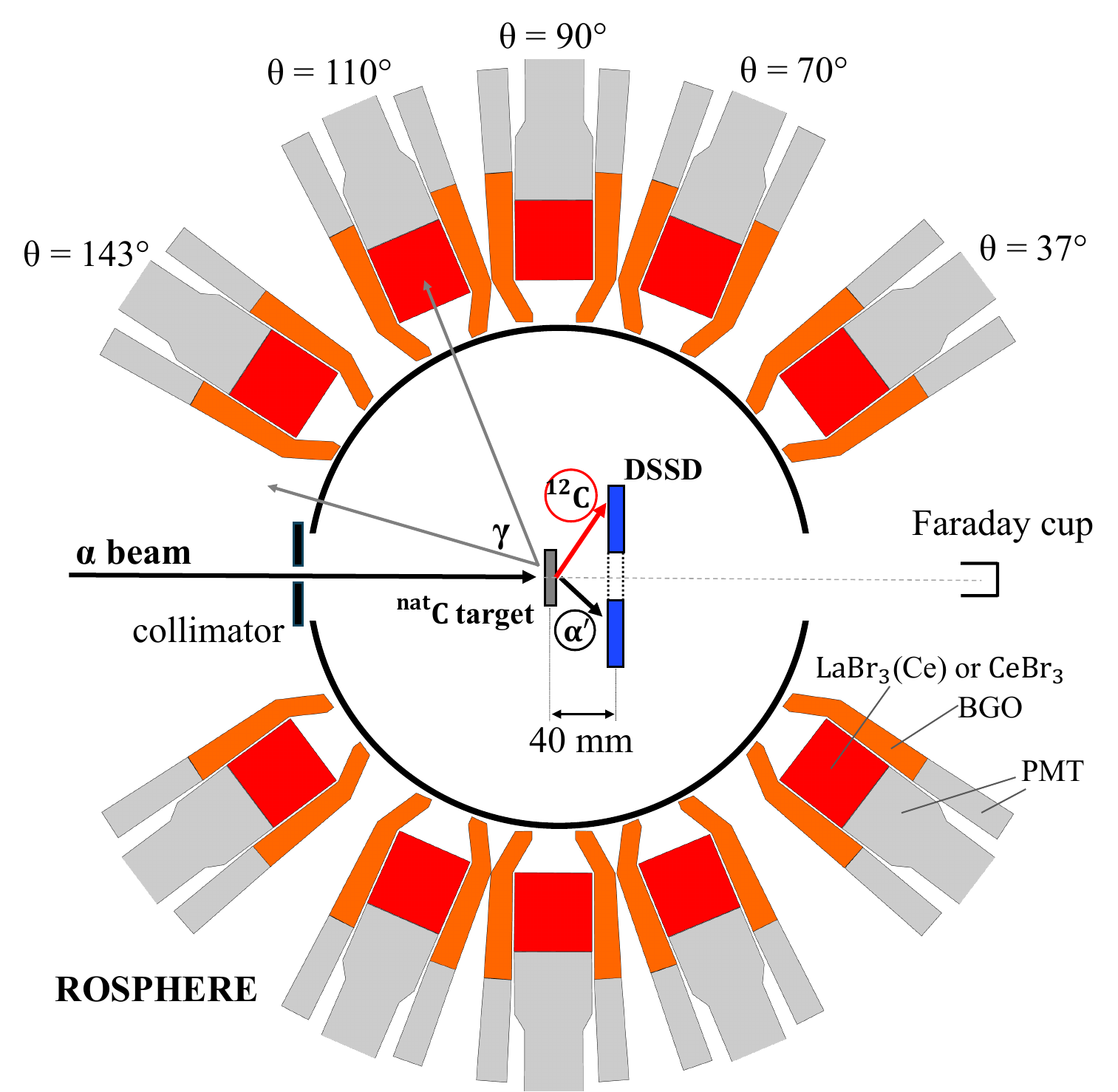}
\caption{Schematic view of the experimental setup.}
\label{fig-2}       
\end{figure}

In this work, we performed a triple-coincidence measurement of the scattered $\rm \alpha$ particles, recoil $\rm ^{12}C$, and emitted $\rm \gamma$ rays to determine the $\gamma$-decay probability of the Hoyle state.
The experiment was conducted at Horia Hulubei National Institute for R\&D in Physics and Nuclear Engineering in Romania.
A $\rm ^{4}He^{2+}$ beam at 25.0 MeV provided by the 9-MV tandem accelerator was transported to the scattering chamber.
Figure \ref{fig-2} shows the experimental setup around the scattering chamber.
The beam impinged on a $\rm ^{nat}C$ target with a thickness of 49 $\rm \mu g/cm^2$ to populate the Hoyle state.
A $\rm ^{13}C$ target with a thickness of 30 $\rm \mu g/cm^2$ and an empty target were also used to subtract background events originated from $\rm ^{13}C$ contained in the $\rm ^{nat}C$ target and from the experimental setup such as a beam-line collimator and a target frame.
The isotopic abundances of the $\rm ^{13}C$ target were 99\% $\rm ^{13}C$ and 1\% $\rm ^{12}C$.
These targets were replaced in cycles of 6 hours, 1 hour, and 0.5 hours, respectively.
Physics data were collected over 90 hours using the $\rm ^{nat}C$ target with an average beam current of 1.5 pnA.
The scattered $\rm \alpha$ particles and recoil $\rm ^{12}C$ were measured with a ring-shaped double sided Si strip detector (DSSD) Design S1 by Micron Semiconductor Ltd.
The inner and outer radii of the sensitive area are 24~mm and 48~mm, respectively.
The thickness of DSSD is 500 $\rm \mu$m.
The junction side of the detector is divided into 4 sector-shape segments with a central angle of 90$^\circ$, and each segment is divided into 16 arc-shaped strips with a pitch of 1.5~mm. 
On the other hand, the ohmic side is separated into 16 radial sectors.
The DSSD was placed at 40 mm downstream of the target to cover $\theta _{\rm lab}$ = $31.0$\textdegree--$50.2$\textdegree.
Signals from the DSSD were processed with Mesytec MPR-16 or MPR-32 preamplifiers, and their pulse shapes were acquired with CAEN V1730SB 14 bit 500 MS/s flash digitizers.
The pulse-shape discrimination (PSD) was employed for particle identification (PID)~\cite{Sakanashi:2025} as described Sec. \ref{sec-PSD}.
The output signals from the preamplifiers were also input to Mesytec MSCF-16 shaping amplifier to generate trigger signals for data acquisition.
Two $\rm \gamma$ rays from the cascade decay of the Hoyle state were detected with the ROSPHERE $\rm \gamma$-ray spectroscopy array consisting of large 14 $\rm LaBr_3(Ce)$ and 10 CeBr$_3$ scintillators.
Each $\rm LaBr_3(Ce)$ and CeBr$_3$ was coupled to a BGO scintillator as the Compton suppressor but the data of the Compton suppressor were not used in the following analysis due to technical issues during the experiment.
Details about ROSPHERE can be found in Refs.~\cite{ROSPHERE:2023,BUCURESCU-2016}.

\section{Analysis}

\subsection{Particle identification with PSD method}
\label{sec-PSD}
In Si semiconductor detectors, a PN junction is utilized to measure radiation.
Since valence electrons intrude from the N-type region near the PN junction surface to the P-type side, a depletion layer is formed with a non-uniform electric field along the depth direction.
When radiation enters the depletion layer, electron-hole pairs are generated.
These pairs are drifted toward readout electrodes by the non-uniform electric field, and induce electric signals on the electrodes.
Since the pair distribution in the depletion layer varies depending on the nuclide and its energy, the signal risetime also changes. 
This change enables PID with the PSD method.

\begin{figure}[htb]
\centering
\includegraphics[width=\linewidth]{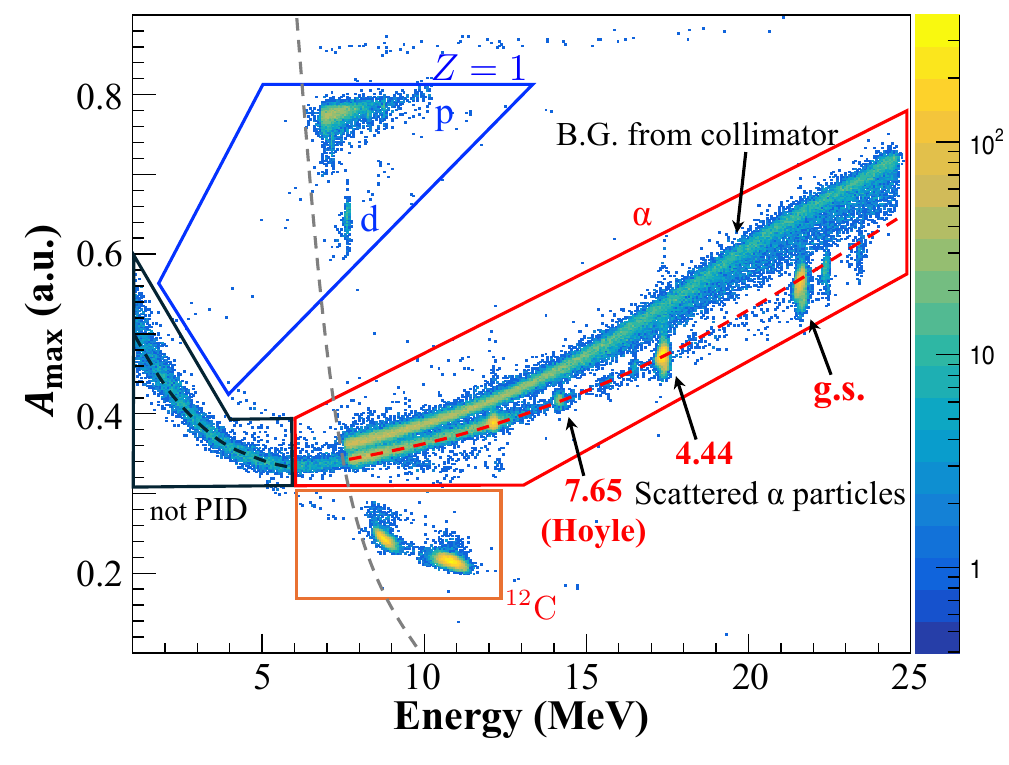}
\caption{Correlation between $A_{\rm max}$ and kinetic energies of detected particles. The red and black dashed lines show the PID function to select $\alpha$-particle-like events. The gray dashed line shows the trigger threshold for the data acquisition.
The red, green, and blue boxes indicate the regions where $\alpha$ particles, $\rm ^{12}C$, and $Z$ = 1 particles (p or d) are located, respectively, while the black box shows the region where PID cannot be performed.
These boxes are drawn as visual guides and do not represent any event-selection conditions in the present analysis.
}
\label{fig-3}       
\end{figure}

The amplitude of the signal obtained by differentiating the output waveform from the preamplifier ($A_{\rm max}$) is a useful PID parameter.
Figure \ref{fig-3} shows the correlation between $A_{\rm max}$ and kinetic energies of detected particles.
We successfully separated scattered $\rm \alpha$ particles from protons, deuterons, and $\rm ^{12}C$ at energies above 6~MeV, although $\alpha$ particles were not distinguishable from $\rm ^{12}C$ at lower energies.
The gray dashed line in Fig. \ref{fig-3} shows the trigger threshold for the data acquisition.
Since the trigger signal was generated with the timing filter amplifier in the MSCF-16 module, the threshold energy varied depending on the risetime of the readout signal from the DSSD.
The low-energy $\alpha$ particles below the trigger threshold in Fig. \ref{fig-3} were decay $\alpha$ particles emitted from excited states in $^{12}$C, which were detected in coincidence with the inelastically scattered $\alpha$ particles.
The continuous locus in Fig. \ref{fig-3} was observed above the discrete loci due to the $\rm \alpha$ particles scattered from the target.
These events in the continuous locus are background particles due to beam particles scattered by beam-line collimators.
Although these background particles were $\alpha$ particles, their $A_{\rm max}$ values showed the different trend from the $\alpha$ particles emitted from the target.
Because the background $\alpha$ particles had smaller incident angles to the DSSD than the $\alpha$ particles from the target, electron-hole-pair distribution along the depth direction of the detector is different and the risetime of the signals was distinguishable.
Only scattered $\rm \alpha$ particles exciting the Hoyle state were selected in the analysis.
The PID functions shown by the dashed red and black lines were determined for each run to estimate the expected $A_{\rm max}$ values for $\alpha$-particle-like events above and below 6~MeV, respectively.
$A^{'}_{\rm max}$  was then defined as the difference between the measured $A_{\rm max}$ value and the expected value from the PID function, so that the $A^{'}_{\rm max}$ values for the $\alpha$-particle-like events became close to zero.

\subsection{{Excitation-energy spectra}}

\begin{figure}[htb]
\centering
\includegraphics[width=8cm]{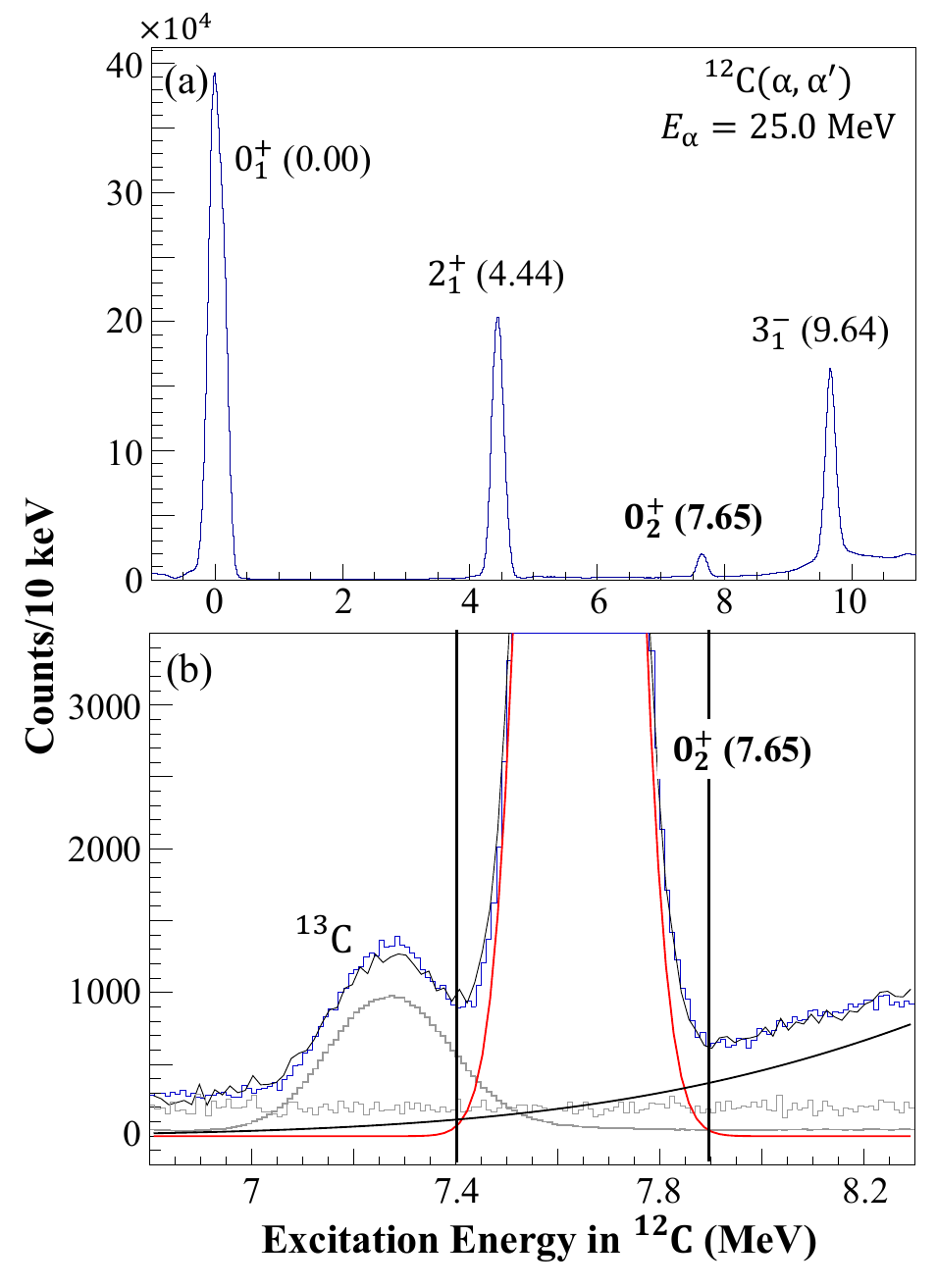}
\caption{(a) Excitation-energy spectrum of $^{12}$C and (b) expanded view around the Hoyle state.
A Gaussian function to fit the Hoyle state is drawn by the thick red line, while an exponential function for high-energy states is shown by the thick black line. Contributions from the $^{13}$C and empty targets are presented by the thick and thin gray lines, respectively.  The thin black line is the sum of the four components.
}
\label{fig-Ex}      
\end{figure}

Figure \ref{fig-Ex}(a) shows the excitation-energy spectrum of $\rm ^{12}C$ with $|A^{'}_{\rm max}|\leq0.025$, while Figure \ref{fig-Ex}(b) presents the expanded view of the spectrum around the Hoyle state. The spectrum was fitted using the experimental spectra obtained with the $^{13}$C target (thick gray line) and the empty target (thin gray line) along with a Gaussian function for the Hoyle state of $^{12}$C (thick red line) and an exponential function for the broad high-energy (H.E.) state including the $2^+_2$ and $0_3^+$ contributions (thick black line).
The thin black line is the sum of the four components.
The scaling factors applied to the experimental spectra of the $^{13}$C and empty targets did not agree with the nominal scaling factor estimated from the irradiated beam intensity and the target thickness, even if their uncertainties were taken into account.
Thus, this inconsistency was considered as a systematic uncertainty in the yield of the Hoyle state $N_\mathrm{H}$.
The value of $N_{\mathrm{H}}$ was obtained by integrating the number of events in the energy range $E_{\mathrm{x}}=7.4$--$7.9$ MeV after subtracting the backgrounds due to the $^{13}$C target, the empty target, and  the H.E. state of $^{12}$C from the experimental spectrum  in Fig. \ref{fig-Ex}(b).
The total number of events in which $^{12}$C was excited to the Hoyle state was determined to be $N_{\mathrm{H}} = (7.63 \pm 0.09) \times 10^6$.
The uncertainty in $N_{\mathrm{H}}$ is mainly systematic, while the statistical uncertainty is negligibly small.

\subsection{$\mathrm{\alpha}$-$\rm ^{12}C$ particle-coincidence method}
\label{sec:a+12C}
First, we determined the radiative-decay probability by detecting the $\alpha$-$^{12}$C particle-coincidence events.
For this purpose, we picked up the coincidence events in which the scattered $\alpha$ particle and the recoil particle were detected within a time window of 30~ns.
To eliminate the background events originated from the 3$\alpha$ decay of the Hoyle state, a coplanarity condition was applied for the event selection.
This condition required that only one recoil particle was detected in the segment on the opposite side of the segment of the DSSD where the scattered $\alpha$ particle hit.
Moreover, the energy of the recoil particle $E_r$ was required to exceed 2.8 MeV in the event selection.
This threshold value was determined by the simulation calculation.
Since the kinetic energy of the recoil $^{12}$C is shared by three $\alpha$ particles in 3$\alpha$-decay events, this requirement completely rejects single-$\alpha$ events, where only one of the three $\alpha$ particles is detected by the DSSD.

\begin{figure}[htb]
\centering
\includegraphics[width=8cm]{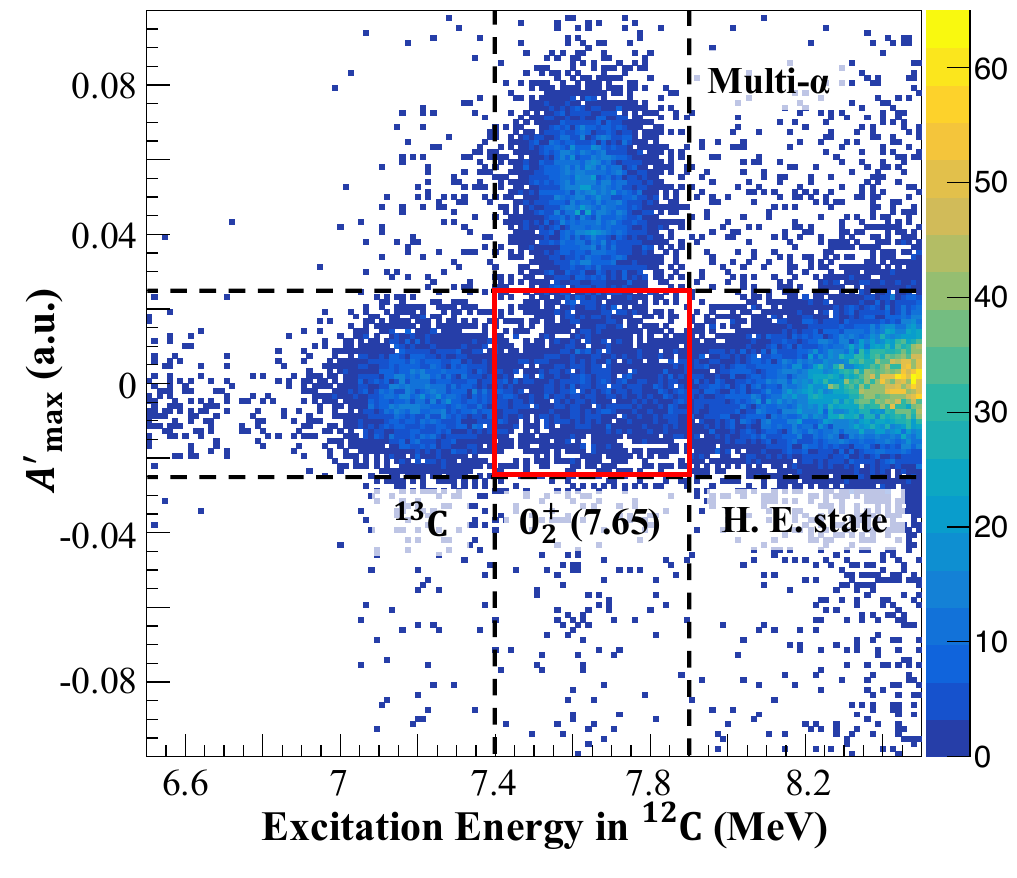}
\caption{Correlation between the excitation energy of $^{12}$C and $A^{'}_{\mathrm{max}}$.
}
\label{fig-ExAmax12C}       
\end{figure}

Figure \ref{fig-ExAmax12C} shows the correlation between the excitation energy of $^{12}$C and $A^{'}_{\mathrm{max}}$ of the recoil particle detected in the candidate $\alpha$-$^{12}$C events.  
Within the red box in the figure, corresponding to the region $E_x$ = 7.40--7.90 MeV and $|A^{'}_{\rm max}|\leq0.025$, 1986 candidate events were observed.  

These candidate events include not only the low-energy tail of the broad H.E. state and $n$+$^{12}$C decay events of excited states in $^{13}$C, but also multi-$\alpha$ events where multiple $\alpha$-particles hit the same matrix of the DSSD.
The detected energies at the matrix could exceed the energy threshold of $E_\mathrm{r}=2.8$ MeV.
Since the readout signal from the matrix is the sum of multiple signals by low-energy $\alpha$ particles, $A^{'}_{\mathrm{max}}$ for the multi-$\alpha$ event is larger than that for the single $\alpha$ particle or $^{12}$C event at the same energy.
Therefore, multi-$\alpha$ events mainly distribute in the large $A^{'}_{\mathrm{max}}$ region in Fig. \ref{fig-ExAmax12C}, but some of the multi-$\alpha$ events could be observed in the signal region shown by the red box.
Since these three types of background could not be removed event-by-event using kinematic and PID conditions, the number of the background events in the red box had to be determined using spectrum fitting.

\begin{figure}[htb]
\centering
\includegraphics[width=8cm]{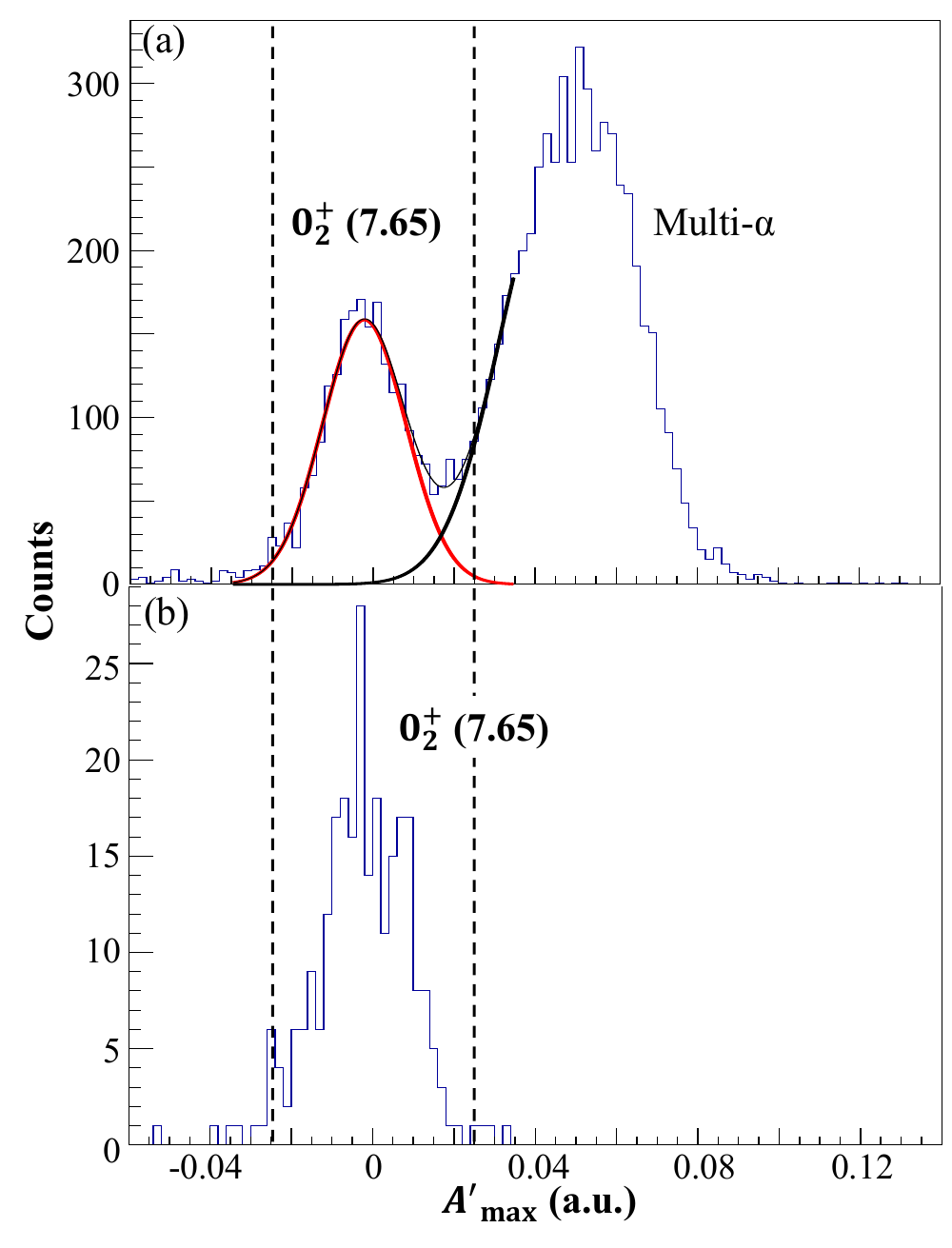}
\caption{(a) Distribution of $A^{'}_{\mathrm{max}}$ for the
  particle-coincidence events at $E_\mathrm{r} \geq2.8$ MeV after
  subtracting accidental coincidence events. The thick red line shows
  a Gaussian function to fit the $^{12}$C events, while the thick
  black line presents an exponential function for the low
  $A^{'}_{\mathrm{max}}$ tail of the multi-$\alpha$ events. The thin
  black line is the sum of the two functions. (b) Distribution of
  $A^{'}_{\mathrm{max}}$ for the triple-coincidence events.
}
\label{fig-ExAmax}       
\end{figure}

Figure \ref{fig-ExAmax}(a) shows the $A^{'}_{\mathrm{max}}$ distribution for the particle coincidence events at $E_x$ = 7.40--7.90 MeV.
Although accidental coincidence events were nearly negligible, they were subtracted in Fig. \ref{fig-ExAmax}(a).
Here, the $^{12}$C events around $A^{'}_{\mathrm{max}}=0$ were fitted with a Gaussian function shown by the thick red line, while the low $A^{'}_{\mathrm{max}}$ tail of the multi-$\alpha$ events was fitted with an exponential function represented by the thick black line.  
Since the actual background distribution is unknown, not only an exponential function but also linear and quadratic functions were employed for fitting.  
The mean and standard deviation of the background contribution were estimated from the results of these fits.  
As the result, the number of multi-$\alpha$ background events within the red box in Fig. \ref{fig-ExAmax}(a) was estimated to be $241\pm16$ (sta.) $\pm$ $39$ (sys.) counts.

\begin{figure}[htb]
\centering
\includegraphics[width=8cm]{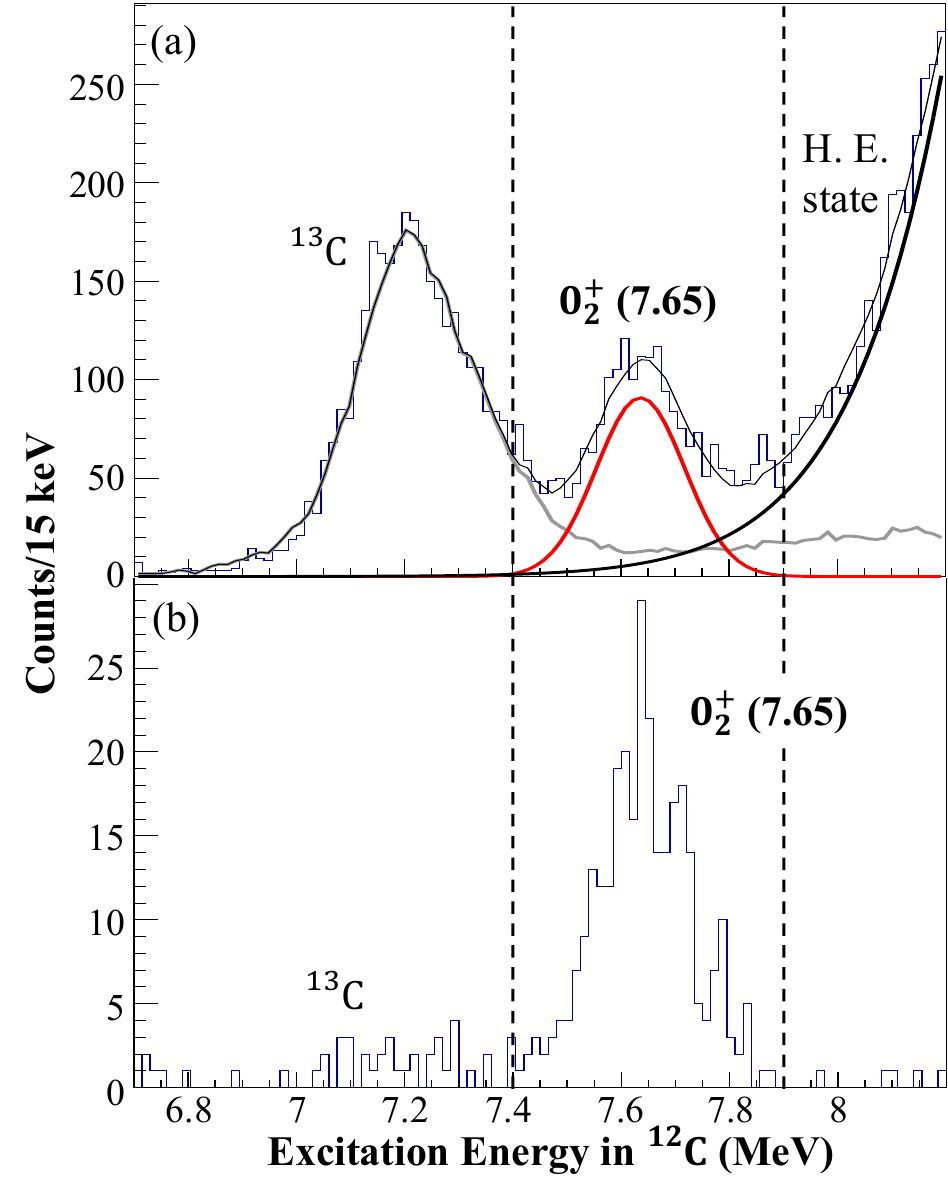}
\caption{(a) Excitation-energy spectrum of the particle-coincidence events with $|A^{'}_{\mathrm{max}}| \leq 0.025$ and $E_\mathrm{r} \geq2.8$ MeV after subtracting accidental coincidence events. The thick red line shows a Gaussian function to fit the Hoyle state, while the thick black line presents a broader Gaussian function for the H. E. state. The thick gray line is the experimental spectrum measured with the $^{13}$C target. The thin black solid line is the sum of three components. (b) Excitation-energy spectrum of the triple-coincidence events.
}
\label{fig-Ex12C}       
\end{figure}

Next, the background contributions from $^{13}$C and the tail of the H. E. energy state of $^{12}$C were evaluated.  
Figure \ref{fig-Ex12C}(a) shows the excitation-energy spectrum of $^{12}$C for the particle-coincidence events with $|A^{'}_{\mathrm{max}}| \leq 0.025$ after subtracting background events caused by accidental coincidences.  
The excitation-energy spectrum was fitted by the experimental spectrum measured with the $^{13}$C target (thick gray line), a Gaussian function corresponding to the Hoyle state (thick red line), and a broader Gaussian function corresponding to the H. E.  state (thick black line).
The thin black line in Fig. \ref{fig-Ex12C}(a) is the sum of the three components.
As the result of the fitting, the number of background events from $^{13}$C and the H.E. state within the red box in Fig. \ref{fig-ExAmax12C} was determined to be $656 \pm 26$ (sta.) and $359 \pm 19$ (sta.) events, respectively.  
Therefore, after subtracting the three background components, the yield of the $\alpha$-$^{12}$C events in the red box of Fig. \ref{fig-ExAmax12C} was determined to be 
$N_\mathrm{^{12}C} = 730 \pm 57 \mathrm{(sta.)} \pm 39 \mathrm{(sys.)}$ counts.

The coincidence-detection efficiency of the $\alpha+{}^{12}$C pairs $\epsilon_{\mathrm{Si}}$ was determined to be $0.219 \pm 0.020$ (sys.) through Monte Carlo simulation taking into account the detector geometry and event selection efficiency with respect to $E_\mathrm{r}$.  
The thickness of the dead layer of the DSSD, which is an important parameter to determine $E_\mathrm{r}$, was evaluated to reproduce the total detected energy of the $\alpha+{}^{12}$C($2^+_1$) coincidence events in the simulation.
The $\alpha+{}^{12}$C($2^+_1$) events are useful to determine the thickness of the dead layer because these events are abundant and the energy of ${}^{12}$C is determined by the energy and angle of the scattered $\alpha$ particle.
The systematic uncertainty of $\epsilon_{\mathrm{Si}}$ was mainly attributed to the uncertainty in the thickness of the dead layer.
Finally, the radiative-decay probability of the Hoyle state was determined to be 
$\Gamma_\mathrm{rad}/\Gamma=[4.3 \pm 0.3 (\rm{sta.}) \pm 0.4 (\rm{sys.})] \times 10^{-4}$ using the $\alpha+{}^{12}$C particle coincidence method.
This value supports the previous study~\cite{Kelley:2017}, however, the large uncertainty prevented the exclusion of either result with sufficient confidence.

\subsection{$\rm \alpha$-$\rm ^{12}C$-$\rm \gamma$ triple-coincidence method}

In the particle-coincidence method, the energy condition on the recoil particles was imposed to reject background events from the 3$\alpha$ decay and to select candidate events for the $\gamma$ decay of the Hoyle state.  
However, this condition introduced a significant uncertainty in the $\mathrm{radiative}$-decay probability because a small variation of the assumed thickness of the dead layer on the DSSD surface led to a considerable change in the evaluated energy of the recoil particle and affected the event selection.  
Additionally, a part of background events caused by the 3$\alpha$ decay and $^{13}$C could not be kinematically eliminated, and they further contributed to the statistical uncertainty in the yield of the $\alpha+{}^{12}$C coincidence events.

On the other hand, since neither of these background sources emits $\gamma$ rays, they can be removed by requiring a $\gamma$-ray coincidence.  
Therefore, in the analysis of $\alpha$-$^{12}$C-$\gamma$ triple-coincidence events within a time window of 30~ns, no energy restriction on the recoil particles was necessary.  
Among the events where the Hoyle state was excited, those that satisfied the coplanarity condition between the scattered and recoil particles, and in which a $\gamma$ ray was detected by ROSPHERE were selected as candidates for $\gamma$-decay events.  

To determine the yield of the triple-coincidence events, the number of accidental events was estimated from the random-coincidence analysis and was found to be less than 10\% of the true coincidence events.

Figures \ref{fig-ExAmax}(b) and \ref{fig-Ex12C}(b) show the distribution of $A^{'}_\mathrm{max}$ and the excitation-energy spectrum for the triple-coincidence events where energy of the detected $\gamma$ ray is greater than 2.6~MeV after subtracting the accidental events.
By applying this $\gamma$-ray coincidence condition, most of the background events that had caused significant uncertainty in the particle-coincidence method were eliminated.

\begin{figure}[htb]
\centering
\includegraphics[width=8cm]{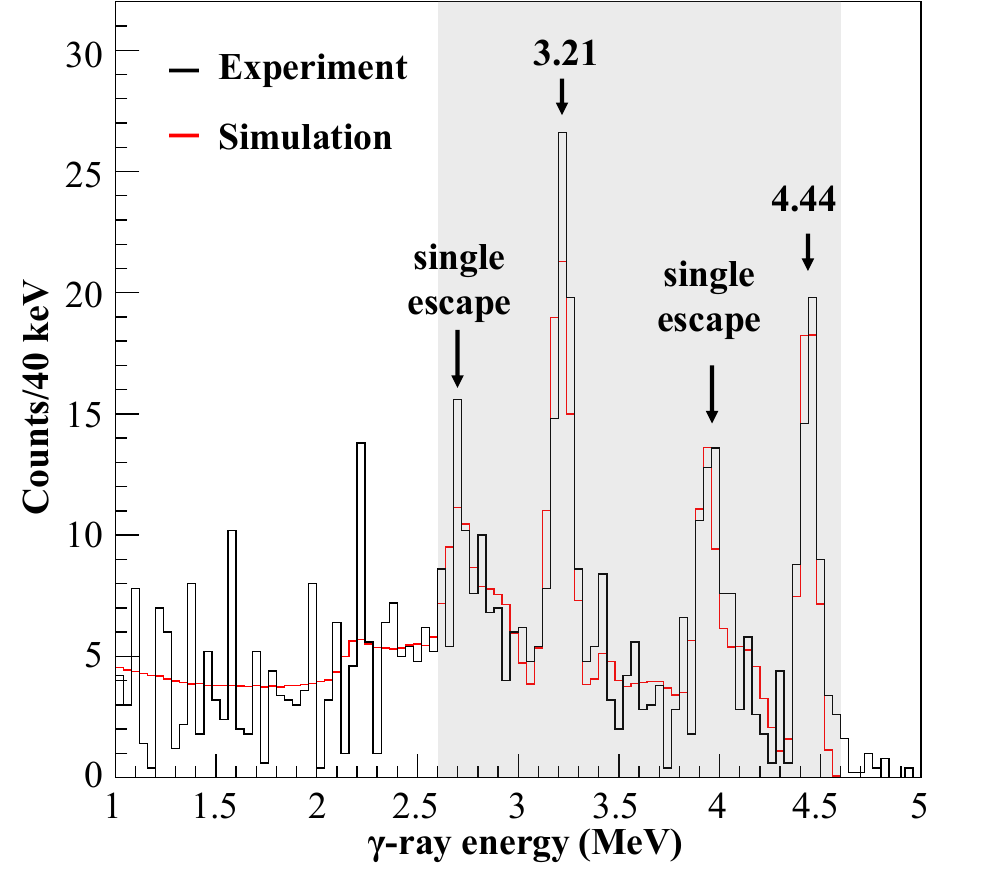}
\caption{$\gamma$-ray energy spectra of the triple-coincidence events at $E_x=7.40$--7.90~MeV from the experimental data (black) and the simulation (red).
}
\label{fig-gamma}       
\end{figure}

The $\gamma$-ray energy spectrum of the triple-coincidence events at $E_x=7.40$--7.90~MeV was obtained after subtracting the accidental coincidence spectrum as shown by the black histogram in Fig. \ref{fig-gamma}.
At $E_\gamma=2.6$--4.6~MeV, the ratio of true to accidental events was 11.
The $\gamma$-ray energy spectrum was fitted using the simulated spectrum as shown in Fig. \ref{fig-gamma}.
The red histogram shows the simulated spectrum.
Since the statistics of the measured $\gamma$ rays in the triple-coincidence events was limited and the uncertainty in each bin was assumed to follow the Poisson distribution, the maximum likelihood method was used to search for the best scaling factor $f$ for the simulated spectrum to fit the experimental spectrum.
In the simulation with Geant4~\cite{geant4}, $N_\mathrm{H}$ events exciting the Hoyle state were generated, and it was assumed that all of them underwent the $\gamma$ decay.  
Thus, 4.44-MeV and 3.21-MeV $\gamma$ rays were assumed to be isotropically generated in equal numbers ($N_\mathrm{H}$) and detected by ROSPHERE.
This assumption is reasonable because the spin and parity of the Hoyle state is $0^+$ and either the 4.44-MeV or 3.21-MeV $\gamma$ ray was detected in the present measurement.

The spectral fitting was performed over the energy range of $E_\gamma = 2.6$--4.6~MeV as indicated by the shaded gray region \highlight{as shown in Fig. \ref{fig-gamma}}, and the scaling factor $f$ was determined to be $f = 3.08 \times 10^4$ with a statistical uncertainty of $0.17 \times 10^4$.
This scaling factor $f$ is related to the number of triple-coincidence events in the measured $\gamma$-ray energy spectrum $N_\gamma$ and the simulated $\gamma$-ray detection efficiency $\epsilon_{\gamma,\mathrm{sim}}$ at $E_\gamma$ = 2.6–4.6 MeV as:

\begin{equation}
N_\gamma = f N_\mathrm{H}\epsilon_{\gamma,\mathrm{sim}}. 
\label{eq:f}
\end{equation}

\noindent 
The maximum likelihood fitting was performed in five different energy regions to estimate the systematic uncertainty originating from inaccuracies in the simulated $\gamma$-ray spectrum, and 
the systematic uncertainty was determined to be $0.05 \times 10^4$ from the standard deviation of the five values. 

The experimental $\gamma$-ray detection efficiency for the 4.44-MeV total-absorption peak $\epsilon^\mathrm{tot}_{\gamma(4.44),\mathrm{exp}}$ was determined using decay events from the first excited ($2^+_1$) state of $^{12}$C, which always decays to the ground state by emitting a 4.44-MeV $\gamma$ ray.  
Since the decay of the $2^+_1$ state is $E2$ transition, the angular distribution of the $\gamma$ rays must be considered to correctly estimate $\epsilon_{\gamma,\mathrm{exp}}$ from the experimental data.
The measured angular distribution was fitted using a function based on Legendre polynomials \cite{PhysRev.131.1260,PhysRev.133.B1368} as:
\begin{equation}
W_\gamma(\theta_{\mathrm{lab}}) 
= 1 + 0.077(58)\times P_2(\cos \theta) - 0.621(12)\times P_4(\cos\theta).
\label{eq:legendre}
\end{equation}

\noindent
Here, $P_2$ and $P_4$ are the Legendre polynomials of degree 2 and 4, respectively.
$\theta_\mathrm{lab}$ is the $\gamma$-ray emission angle with respect to the beam axis in the laboratory frame.
We adopted the same method as that used in Refs. \cite{Alburger:1977,Eriksen:2020}.
As the result, the total-absorption efficiency of ROSPHERE for a  4.44-MeV $\gamma$ ray isotropically emitted from the target was determined to be $\epsilon^\mathrm{tot}_{\gamma(4.44),\mathrm{exp}} = 2.41(9)\%$, and the ratio of experimental to simulated efficiency $\epsilon^{\mathrm{tot}}_{\gamma (4.44), \mathrm{exp}} /
\epsilon^{\mathrm{tot}}_{\gamma (4.44), \mathrm{sim}}$ was determined to be $0.91(3)$.  
The uncertainty of $\epsilon^{\mathrm{tot}}_{\gamma(4.44),\mathrm{exp}}/
\epsilon^{\mathrm{tot}}_{\gamma(4.44),\mathrm{sim}}$ is mainly systematic, originating from the fitting analysis of the angular distribution of the $\gamma$ ray, while the statistical uncertainty was negligible due to the large population of the $2^+_1$ state.
The coincidence-detection efficiency of the $\alpha+{}^{12}$C pair by the DSSD was estimated to be $\epsilon_{\mathrm{Si}}=0.846(16)$ from the detector configuration.
In this estimation, systematic uncertainties arising from the shifts in the position of the beam spot on the target monitored during the experiment, as well as the ambiguity in the distance between the target and the DSSD were considered.
It should be noted that this value of $\epsilon_\mathrm{Si}$ is different from that used for the $\alpha+{}^{12}$C particle-coincidence method in Sec. \ref{sec:a+12C} as no event selection based on the recoil-particle energy was applied here.

Finally, the $\gamma$-decay probability $\Gamma_\gamma/\Gamma$ was determined as:
\begin{equation}
\begin{aligned}
\frac{\Gamma_\gamma}{\Gamma} &= \frac{N_\mathrm{\gamma}}{N_\mathrm{H}} 
\times \frac{1}{\epsilon_\mathrm{Si}} \times \frac{1}{\epsilon_\mathrm{\gamma,exp}}, \\
&=\frac{N_\gamma}{N_H}\times\frac{1}{\epsilon_\mathrm{Si}}\times\frac{1}{\epsilon_{\gamma,\mathrm{sim}}}\frac{\epsilon_{\gamma,\mathrm{sim}}}{\epsilon_{\gamma,\mathrm{exp}}}. 
\end{aligned}
\label{eq:pro_f}
\end{equation}

\noindent
Using Eqs. (\ref{eq:f}) and (\ref{eq:pro_f}):
\begin{equation}
\begin{aligned}
\frac{\Gamma_\gamma}{\Gamma}
&= f\times \frac{1}{\epsilon_\mathrm{Si}}
\frac{\epsilon_{\gamma,\mathrm{sim}}}
{\epsilon_{\gamma,\mathrm{exp}}}.
\end{aligned}
\label{eq:pro_ff}
\end{equation}

The correction factor for the simulated $\gamma$-detection efficiency $\epsilon_{\gamma,\mathrm{sim}}/\epsilon_{\gamma,\mathrm{exp}}$ could be approximated by that for the total-absorption efficiency for the 4.44-MeV $\gamma$ ray $\epsilon^{\mathrm{tot}}_{\gamma (4.44), \mathrm{exp}} / \epsilon^{\mathrm{tot}}_{\gamma (4.44), \mathrm{sim}}$.
Finally, the $\gamma$-decay probability was obtained as $\Gamma_\gamma/\Gamma=[4.00 \pm 0.22 (\rm{sta.}) \pm 0.18 (\rm{sys.})] \times 10^{-4}$

\section{Discussion}
\begin{figure}[htb]
\centering
\includegraphics[width=8.5cm]{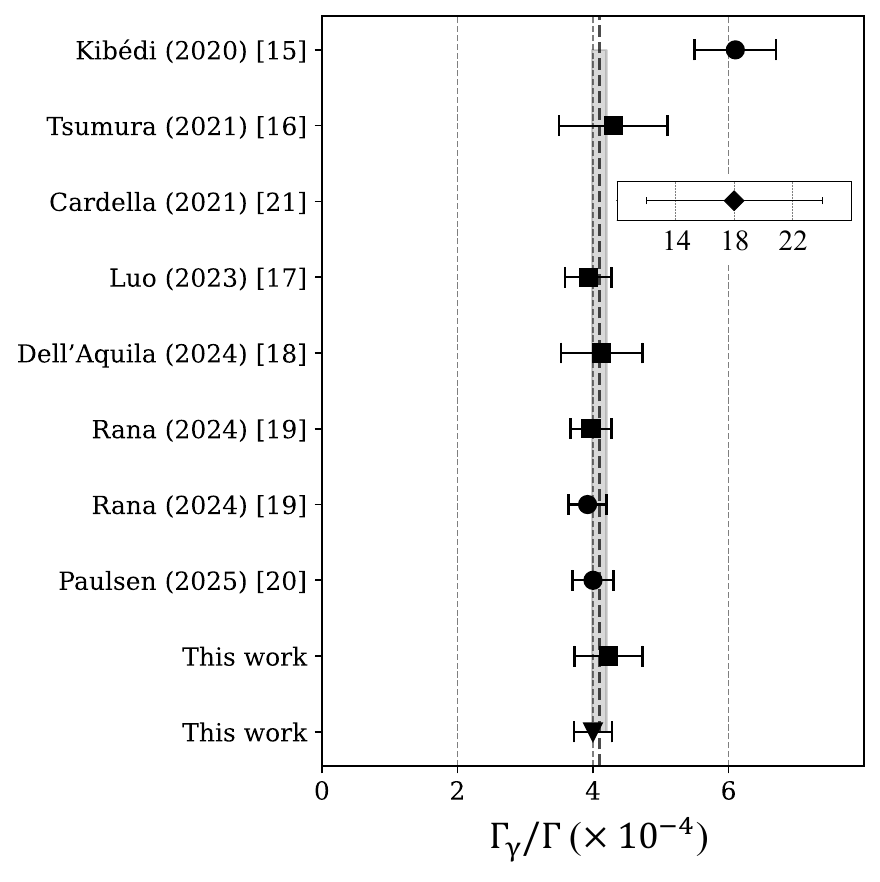}
\caption{Summary of the $\gamma$-decay probability reported after
the study by Kibédi $et$ $al.$~\cite{kibedi2020}. The solid squares and circles present the results obtained from the measurements of surviving $^{12}$C and decay $\gamma$ rays, respectively. The solid triangle shows the present result obtained by the $\alpha$-$^{12}$C-$\gamma$ triple-coincidence method, while the solid diamond in the inset indicates the result by the quadruple-coincidence method.}
\label{fig-result}       
\end{figure}

Various experimental approaches have been employed to determine the radiative-decay probability of the Hoyle state after Kib\'edi $et$ $al.$ as summarized in Fig. \ref{fig-result}.
The solid squares and circles present the results obtained from the measurements of surviving 
$^{12}$C~\cite{{tsumura,newpaper2,newpaper3,newpaper4}}, and decay $\gamma$ rays~\cite{newpaper4,Paulsen:2025}, respectively.
All these studies with the two different methods are consistent with the previous literature value adapted in Ref.~\cite{Kelley:2017}, but contradicts the new value by Kib\'edi $et$ $al.$~\cite{kibedi2020}.
Cardella $et$ $al.$ conducted a measurement combining these two methods~\cite{newpaper1}. 
They measured all particles emitted from inelastic $\alpha$ scattering that excited the Hoyle state and its subsequent decay in quadruple coincidence involving the scattered $\alpha$ particle, surviving $^{12}$C, and two decay $\gamma$ rays.
They reported the $\gamma$-decay probability approximately 4.5 times larger than the literature value as indicated by the solid diamond in the inset of Fig. \ref{fig-result}.
This value is even about 3 times higher than the result reported by Kibédi $et$ $al.$
As a possible explanation for this large value, the existence of an Efimov state~\cite{efimov} near the Hoyle state with a large $\gamma$-decay probability was proposed~\cite{Cardella:2023}.
However, if this hypothesis were correct, a similar enhancement of the $\gamma$-decay probability should be observed in the other experiments.
The $\gamma$-decay probability reported by Cardella $et$ $al.$ was possibly overestimated due to an incorrect evaluation of the background events caused by the accidental coincidences.

In the present work, we successfully achieved low background levels using a novel approach with the $\alpha$-$^{12}$C-$\gamma$ triple-coincidence method.
The use of a DC beam provided by the tandem accelerator significantly suppressed accidental coincidence events.
The present work yielded the reliable result supporting the previous literature value as shown by the solid triangle in Fig. \ref{fig-result}.
Thus, we conclude that the literature value of the $\gamma$-decay probability adopted in Ref.~\cite{Kelley:2017} can be reliably used to calculate the 3$\alpha$ reaction rate.

\section{Summary}
In summary, we measured the $\gamma$-decay probability of the Hoyle state with a new method of triple coincidence detection of a scattered $\alpha$ particle, a recoil $\rm ^{12}C$ nucleus, and a $\gamma$ ray in the inelastic alpha scattering on  $\rm ^{12}C$.
This method enabled low background levels, and successfully determined the $\gamma$-decay probability of the Hoyle state as $\Gamma_\mathrm{\gamma}/\Gamma=[4.00 \pm 0.22 \mathrm{(sta.)} \pm 0.18 \mathrm{(sys.)}]\times10^{-4}$.
By adopting the value of the pair-decay branching ratio $\Gamma_\pi/\Gamma=(6.7 \pm 0.6)\times10^{-6}$ from Ref. \cite{Kelley:2017}, the radiative-decay probability was deduced to be $\Gamma_\mathrm{rad}/\Gamma=4.07(28)\times10^{-4}$.
This result supports the previous literature value as shown by the solid triangle in Fig. \ref{fig-result}.
Thus, we conclude that the literature value of the $\gamma$-decay probability adopted in Ref.~\cite{Kelley:2017} can be reliably used to calculate the 3$\alpha$ reaction rate.

\section*{Acknowledgements}
K. Sakanashi appreciates the support from the JSPS Research Fellowships for Young Scientists. This work was partly supported by JSPS KAKENHI Grants No. JP19H05604 and No. JP22K14059, and Bilateral Collaboration No. JPJSBP120239922.

This work was also partially supported by the ELI-RO
program funded by the Institute of Atomic Physics, M\u{a}gurele,
Romania, contract number ELI-RO/RDI/2024-002 (CIPHERS) and the Romanian
Ministry of Research and Innovation under research contract PN~23~21~01~06.

\bibliographystyle{elsarticle-harv} 
\bibliography{ref}

\end{document}